**Computational modeling to elucidate molecular mechanisms of epigenetic memory**

Jianhua Xing[1, 2*], Jin Yu[1], Hang Zhang[1,3], Xiao-Jun Tian[2]

[1]Beijing Computational Science Research Center, Beijing 100084, China

[2]Department of Computational and Systems Biology, School of Medicine, University of Pittsburgh, Pittsburgh, Pennsylvania, 15213, USA

[3]Department of Biological Sciences, Virginia Polytechnic Institute and State University, Blacksburg, Virginia, 24060, USA

[*] Correspondence should be sent to xing1@pitt.edu

Abstract (100-250 words)

How do mammalian cells that share the same genome exist in notably distinct phenotypes, exhibiting differences in morphology, gene expression patterns, and epigenetic chromatin statuses? Furthermore how do cells of different phenotypes differentiate reproducibly from a single fertilized egg? These are fundamental problems in developmental biology. Epigenetic histone modifications play an important role in the maintenance of different cell phenotypes. The exact molecular mechanism for inheritance of the modification patterns over cell generations remains elusive. The complexity comes partly from the number of molecular species and the broad time scales involved. In recent years mathematical modeling has made significant contributions on elucidating the molecular mechanisms of DNA methylation and histone covalent modification inheritance. We will pedagogically introduce the typical procedure and some technical details of performing a mathematical modeling study, and discuss future developments.

Keywords (5-10)

Histone epigenetic memory, mathematical modeling, reader and writer, pattern reconstruction, stochastic dynamics, histone modification enzyme, post-translational modification

Outline:

In this chapter we discuss how to use mathematical modeling to explore the dynamics and mechanism of histone modification dynamics.

Textbody

## 1. Introduction

There are more than 200 different cell types in a human body. These cells have drastically different shapes, physical and physiological properties. Amazingly, all these cells (except reproductive cells) share the same set of genomes, and are developed from a single fertilized egg. Therefore, a fundamental and intriguing question in developmental biology is how a fertilized egg can develop into so many different types, in a controlled manner. Furthermore, a cell can preserve its identity after division. That is, a fibroblast cell divides into fibroblast cells. Recent studies show that it is possible, but difficult to reprogram the identity of a terminally differentiated cell (1). Then how can a cell remember its identity? Nowadays accumulating evidences suggest that some heritable changes of gene activities are not caused by changes in the DNA sequence. Specifically in this chapter we will focus on heritable histone covalent modifications.

To form an organized and compact chromatin structure, a DNA molecule wraps around histone octamers to form nucleosomes. Covalent modifications such as methylation, acetylation, phosphorylation and sumoylation can take place on a number of side residues of the histone proteins. Through changing the interactions between DNA and histone proteins, and between nucleosome and other regulatory elements such as histone modification enzymes, transcription factors and regulatory noncoding RNAs, these covalent modifications affect higher-order packing of the nucleosomes and gene transcription efficiency (2). Experimental studies reveal that at least some of the histone post-translational modification patterns are inheritable, which is called histone epigenetics memory (3, 4). In the past few years, different groups had discovered multiple enzymes regulating the histone modification dynamics (3, 5). The so-called 'histone code' proposal, although still under debate, has drawn extensive attention from the field (6).

Revealing the molecular mechanism of this histone modification memory has become a focused research subject for many years.

In recent years mathematical modeling has contributed significantly to our understanding how histone epigenetic patterns are produced and maintained. In a seminal paper, Dodd *et al.* used a rule-based model to analyze the silenced mating-type locus of the fission yeast *Schizosaccharomyces pombe* (*S. Pombe*) (7). *S. Pombe* has two mating-type cassettes that are normally in an epigenetically silent state. A mutant has been constructed with portion of the silenced region removed and an ura4+ reporter gene inserted. Experimental studies on the mutant revealed that the DNA region (~ 60 nucleosomes) can exist in an inheritable epigenetic active or silent state, with a very low probability of stochastic transition between the two states of about $5 \times 10^{-4}$ per cell division (8, 9). Furthermore, the two copies of the chromosomal region within one cell can exist in different epigenetic states. That is, cells can exist in a bistable region. The mathematical analysis of Dodd *et al.* showed that cooperativity among neighboring and beyond-neighboring nucleosomes are necessary and sufficient to generate robust bistable epigenetic states. Subsequently this pioneering study has been generalized to analyze systems such as vernalization in Arabidopsis Thaliana (10), epigenetic switching at the genetic locus of Oct4 (also known as Pou5f1), a transcription factor essential for maintaining the embryonic stem cell state (11, 12), and olfactory neuron differentiation (13). Meanwhile studies using alternative approaches have also been developed to analyze various problems (14-24). Especially, quantitative measurements on nucleosome covalent modification dynamics allow incorporation of molecular details in modeling studies. Steffen *et al.* (25) and Rohlf *et al.* (26) provided nice and timely reviews on the experimental and mathematical modeling efforts to extract quantitative information of epigenetic regulation. In the remaining of the chapter, we will discuss in detail the generic procedure of performing a mathematical modeling study. We will use a model of Zhang *et al.*, which has all its components based on experimental information (27), as an example. The model has its structure similar to the well-studied Potts model in physics describing cooperative phenomenon. For simplicity we will call it the CoPE model, standing for coupled-Potts model of epigenetic dynamics of histone modifications.

## 2. Identify puzzle from experimental studies

The first step for a modeling study is to identify a problem that is both significant and suitable for theoretical studies. Modeling is not intended and is not capable of answering every question. For example, modeling studies can examine whether a proposed mechanism is consistent with available experimental observations, and the laws of physics and chemistry, but cannot decide whether the mechanism is actually what assumed by the system. The confirmation must come from experimental studies. Similarly modeling studies may suggest whether a missing component is needed to reconcile existing data, but cannot determine the identity of the component.

For information inheritance from mother to daughter cells, the puzzle is how the information is transmitted and maintained. We can identify three types of heritable information: the DNA, whose double helix structure allows faithful reproduction; the abundance of proteins and other molecules (i.e., the transcriptome, proteome, *etc.*), which partition into two daughter cells either equally or asymmetrically; the covalent modification patterns on DNA molecules and on histones, whose inheritance mechanisms are less understood. For concreteness in this chapter we will focus on the problem of histone pattern inheritance, while the procedure can be easily generalized to DNA methylation as well.

A closer examination of the histone inheritance problem reveals that it is a highly nontrivial question. First, within a nucleus, there are constantly opposing histone modification enzymes attempting to add or remove each covalent mark and modify the histone pattern; thus instead of being static, the histone modification pattern is a consequence of dynamic "tug-of-war". Second, although the interactions between a histone complex and DNA are not weak, the histone can stochastically detach from the DNA, then either the same histone or a new one, which likely bears no or covalent mark(s) different from the old one, quickly incorporates onto the DNA. This process is termed histone turnover. Furthermore, when cell division takes place, each histone is likely partitioned into one of the two daughter cells with equal probability; that is, each daughter cell needs to incorporate about half of the total DNA-binding histones with nascent unmarked histones. Amazingly, with all these large perturbations, cells can maintain at least some of histone covalent patterns for generations.

Extensive biochemistry and biophysics studies reveal two prominent properties of the modification enzymes. First, the enzymes can recognize histone marks and thus have different free energy of binding. For example, Jacobs and Khorasanizadeh reported the structural basis for the chromodomain of *Drosophila* HP1 to recognize the trimethylated H3K9 residue (28), Raphael Margueron *et al.* discovered that H3K27me3 propagation and maintenance require specific recognition of H3K27me3 by polycomb protein complex (29). Generally speaking, an enzyme has higher binding affinity to nucleosomes bearing the corresponding marks than those without mark or with different marks (30). We want to point out that this property is typical for enzymes, *i.e.*, an enzyme usually binds stronger to the substrate than to the product or to a non-substrate. Second, enzymes bound to neighboring nucleosomes can interact laterally. Canzio *et al.* showed that the HP1 proteins can form oligomers through chromodomain and chromoshadow domain lateral interactions, and enhanced lateral interactions lead to higher percentage of H3K9me3 (31, 32). Interestingly, mutations related to the histone modification enzyme lateral interactions have been reported in cancer cells (33).

Therefore the puzzle, or the question we want to address is whether one can use the above-discussed molecular level information to explain the epigenetic histone memory. The process is complex, with many molecular species, and broad time scales involved. For the latter it ranges from subsecond for enzyme binding/unbinding events, to months or longer for histone memory duration. For example, epigenetic state switches for the above-mentioned *S. Pombe* mutant take place about every 200 days on average (8, 9). Therefore mathematical modeling is necessary to fill in the huge gaps between the experimentally explored molecular events and collective epigenetic dynamics.

### 3. Formulate mathematical model

With the problem identified, next one needs to translate it into a mathematical model. Here we use the word "translate" literally. That is, each term in the mathematical model corresponds to a process identified as important for understanding histone memory. One does, however, need to consider carefully on what levels of details to be included. In physics, a common criterion is based on the following famous quote from Einstein, a theory "should be made as simple as possible, but not simpler". That is, the model should contain just the right amount of details sufficient to explain the underlying phenomenon, but not more to distract one from the essential physics. For example, if one only wants to know the dimension of a box, then information about the box color is irrelevant. To keep a model necessarily simple, abstraction is often needed.

**Insert Figure 1**

Figure 1 summarizes the CoPE model, which includes a collection of *N* nucleosomes aligned as a one-dimensional array. A nucleosome *i* has three possible covalent states, bearing repressive mark, unmarked, or bearing active mark. For bookkeeping purpose, let's denote them as $s_i$ = -1, 0, 1, respectively. In addition four classes of covalent modification enzymes can bind to each nucleosome to catalyze adding or removing the marks. Thus each nucleosome can have 5 possible enzyme binding states, empty or one type of the enzymes bound, which we denote as $\sigma_i$ (= 1-5), indicating no enzyme bound ($\sigma_i$ = 1), repressive modification addition enzyme bound ($\sigma_i$ = 2), repressive modification removal enzyme bound ($\sigma_i$ = 3), active modification addition enzyme bound ($\sigma_i = 4$), active modification removal enzyme bound ($\sigma_i = 5$). The σ state can change through enzyme binding and unbinding. The state of the system is thus denoted by the set of nucleosome indices $\{s_i\, \sigma_i,\, i = 1, \ldots, N\}$.

The overall *s-σ* state of the system evolves according to a Markovian dynamics. That is, the evolution depends only on the state in a previous time step. Enzyme binding/unbinding results in the *σ*-state change. The *s* state can change through histone turnover or enzyme catalyzed chemical reactions. For the latter each of such reactions clearly requires that the corresponding enzyme binds to the nucleosome. Another relevant process is cell division. After each cell division, histones from a mother cell partition into two daughter cells. Current evidences suggest that this partition is random with equal probability to the daughter cells. Then nascent unmarked histones need to be incorporated to the DNA. In the language of modeling, for any given DNA-bound histone, during cell division its *s* state is randomly decided to either keep its current value or reset to 0 with equal probability.

The covalent modification enzymes have no DNA sequence specificity. That is, they do not know which genome region to modify. Accumulating evidences suggest that some regulatory elements, such as transcription factors and non-coding RNAs, may recruit certain enzymes to specific DNA regions (34). For example, the transcription factor SNAIL1 recruits to the E-cadherin promoter region histone demethylase LSD1 that removes H3K4me2 (35), histone deacetylase 1 (HDAC1) and HDAC2 (36), and PRC2, an H3K27me3 methyltransferase (37). In addition, some enzymes, *e.g.* MLL1, KDM2A, PRC2 have higher binding affinity at some DNA sequence elements, *e.g.*, CpG islands (38-41). To reflect these observations, we follow the treatment of Angel *et al.* (10), and Hodge and Crabtree (11), to denote a "nucleation region" for a small number of nucleosomes, on which the enzymes have higher binding affinity compared to the nonspecific background binding affinity on other nucleosomes. Existence of the nucleation regions can be inferred from the peaked distribution of histone modifications centered around many transcription factor binding sites (42).

Clearly the model is rather generic, and has neglected lots of details. Below we just list a few.

1) Many residues can exist in multiple modification states. For example, a lysine can be mono-, di-, and tri-methylated, with different enzymes catalyze each methylation and demythelation step. Also covalent modification enzymes may function redundantly and act on different substrates. For example, LSD1 (*lysine-specific demethylase 1*) can remove both mono- and di-methylation on H3K9 and H3K4. Both PRC1 (*Polycomb Repressive Complex 1*) and PRC2 (*Polycomb Repressive Complex 2*) can catalyze trimethylation on H3K27.

2) Each histone can have a large number of potential modification sites, leading to an even larger combinatory number of epigenetic states. According to the epigenetic code hypothesis, the covalent states of some sites may mutually affect each other and lead to different regulation on the gene activity,

3) A histone modification enzyme complex is usually bulky, and can interact with more than one nucleosome simultaneously.

4) The three dimensional structure of chromatin affects the histone modification dynamics, *e.g.*, accessibility to the enzymes. In return, histone modifications may affect the three-dimensional packing of the chromatin.

These details likely have various biological implications. It is straightforward to expand the CoPE model to incorporate these details. However, the main purpose of that work is to uncover the most essential molecular interactions and properties for histone memory. Therefore, these complexities are not explicitly considered. As we emphasized above, simplification is a key step for modeling.

### 4. Choose appropriate modeling techniques

The above-mentioned model is straightforward in terms of describing the relevant biological processes. However, technical difficulties exist on studying it. Even with this simplified model, each nucleosome has 3 $s$ states and 5 $\sigma$ states. With $N$ nucleosomes, the total number of states is $15^N$, which grows quickly with $N$. Furthermore, the possible dynamic processes, including enzyme binding and unbinding, chemical reactions, histone turnover, and cell cycles, span broad time scales, from sub-second binding/unbinding events to the epigenetic state switching on the order of days to years. This large number of states and the broad time scale distribution make it computationally very expensive to simulate the system. Fortunately the time scale of enzyme binding/unbinding is well separated from that of other processes, which suggests a quasi- equilibrium approximation.

One may remember the quasi-equilibrium approximation on deriving the Michaelis-Menten equation for enzymatic dynamics. One assumes that an enzymatic reaction follows the following scheme,

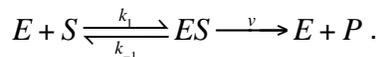

$$E + S \underset{k_{-1}}{\overset{k_1}{\rightleftharpoons}} ES \overset{v}{\longrightarrow} E + P \ .$$

That is, enzyme $E$ and substrate $S$ first form a complex $ES$, which then proceeds to form the product $P$ and release the enzyme for the next enzymatic cycle. The quasi-equilibrium approximation assumes that the first step of forming $ES$ is fast compared to the covalent bond breaking/forming step, so that $E$, $S$ and $ES$ concentrations reach an equilibrium distribution, $\frac{[ES]}{[E]} = \frac{k_1[S]}{k_{-1}} = e^{-\varepsilon/(k_B T)}$, where $\varepsilon$ is the free energy of $S$ binding to $E$ at concentration $[S]$, (notice that $-\varepsilon$ is the binding affinity), $k_B$ is the Boltzmann's constant, $T$ the temperature, and 1 $k_B T$ is ~ 0.6 kcal/mol at room temperature. If the total enzyme concentration is conserved, $[E] + [ES] = [E]_{tot}$, one has

$$[ES] = \frac{k_1[S]}{k_{-1}+k_1[S]}[E]_{tot} = \frac{e^{-\varepsilon/(k_BT)}}{1+e^{-\varepsilon/(k_BT)}}[E]_{tot}. \quad \text{Eqn. 1}$$

Here for convenience of the following discussions, we have written the above expression in the form of the Boltzmann distribution. Notice that an enzyme molecule can exist either in a free form *E*, or a bound form *ES*. If we set the *E* state, which we number as state 1, with a free energy level $\epsilon_1 = 0$, then the *ES* state, which we number as state 2, has a free energy level $\epsilon_2 = \varepsilon$. The Boltzmann distribution states that the probability of finding an enzyme in *ES* state is given by

$$p_1 = \frac{1}{Z}, p_2 = \frac{e^{-\varepsilon/(k_BT)}}{Z}, \quad \text{Eqn. 2}$$

where $Z = \sum_{i=1}^{2} e^{-\epsilon_i/(k_BT)} = 1 + e^{-\varepsilon/(k_BT)}$ is called the partition function in statistical physics, and is defined as summation over the Boltzmann factors of all states. Then from $[ES] = p_2[E]_{tot}$, one recovers Eqn. 1. With Eqn. 1 one obtains the familiar Michaelis-Menten kinetics equation (under the quasi-equilibrium approximation),

$$\frac{d[P]}{dt} = k_2[ES] = \frac{k_1v[S]}{k_{-1}+k_1[S]}[E]_{tot} = \frac{ve^{-\varepsilon/(k_BT)}}{1+e^{-\varepsilon/(k_BT)}}[E]_{tot} = vp_2[E]_{tot}. \quad \text{Eqn. 3}$$

In the CoPE model, Zhang *et al.* adopts a similar approximation, although it is a little more complicated since enzymes can bind on any of the *N* nucleosomes and catalyze chemical reactions. For simplicity, let's consider a case with two nucleosomes. There are 9 possible covalent states specified by {$s_1$, $s_2$}. For each of them, there are 25 possible enzyme binding states specified by {$\sigma_1$, $\sigma_2$}. Again we can assign each enzyme binding state a free energy level $\epsilon_{s_1s_2;\sigma_1\sigma_2} = \varepsilon_{s_1\sigma_1} + \varepsilon_{s_1\sigma_1} - J_{\sigma_1\sigma_2}$. Notice that the free energy of binding $\varepsilon$ is *s*-dependent, and a term $-J_{\sigma_1\sigma_2}$ represents the lateral interactions between two enzymes bound to the two neighboring nucleosomes. The Boltzmann distribution gives the probability of finding the system, *i.e.*, the two nucleosomes in state {$s_1$, $s_2$; $\sigma_1$, $\sigma_2$} is $p_{s_1s_2;\sigma_1\sigma_2} = e^{-\epsilon_{s_1s_2;\sigma_1\sigma_2}/(k_BT)}/Z_{s_1s_2}$. The partition function $Z_{s_1 s_2}$ is obtained by summing over all the 25 enzyme binding states with fixed covalent state {$s_1$, $s_2$}. The expression, $\bar{p}_{\sigma_1} = \sum_{\sigma_2=1}^{5} e^{-\frac{\epsilon_{s_1 s_2; \sigma_1 \sigma_2}}{k_B T}}/Z_{s_1 s_2}$, gives the probability of finding one nucleosome, *e.g.*, nucleosome 1, with enzyme binding state, $\sigma_1$, irrespective of the enzyme binding state of nucleosome 2. Then we can obtain the enzymatic reaction rate for a specific reaction using an expression similar to Eqn. 3. For example, suppose that the two nucleosomes are in state {$s_1$ = -1, $s_2$ = 0}. Nucleosome 1 can have its repressive mark removed by a bound corresponding removal enzyme $E_{Rr}$, and the rate is given by $k_1 = v_{-1 \to 0} \bar{p}_{\sigma_1 = 3}[E_{Rr}]_{eff} = v'_{-1 \to 0} \bar{p}_{\sigma_1 = 3}$, where $[E_{Rr}]_{eff}$ is the effective repressive mark removal enzyme concentration in the nucleus. With the quasi-equilibrium approximation, we separate the $3^N$ *s* states and $5^N$ *σ* states, and remove the necessity of treating the binding/unbinding processes explicitly, thus greatly reduce the computational cost.

With the enzyme binding/unbinding processes treated by the above quasi-equilibrium approximation, the following events can take place:

1) An enzymatic reaction or a process of histone turnover at site *i* with rate $k_i = \delta_{s_i,0}(v'_{0 \to -1}\bar{p}_2 + v'_{0 \to 1}\bar{p}_4) + \delta_{s_i,-1} v'_{-1 \to 0}(\bar{p}_3 + d) + \delta_{s_i,1} v'_{1 \to 0}(\bar{p}_5 + d)$. Here $\delta_{ij}$ is the Kronecker delta function, which assumes a value 1 when *i = j*, and 0 when *i ≠ j*. Notice here we take into account the fact that for an enzymatic reaction to take place, the corresponding enzyme has to bind to the nucleosome. The term *d* is the histone replacement rate due to stochastic turnover ($s_i \to 0$).

2) Every time when cell division takes place, each histone has 50% probability to be partitioned to one of the daughter cells.

Therefore the overall simulation procedure is as follows:

For each step with covalent state {$s_i$},

1) Calculate $\bar{p}_{\sigma_i}$ and $\{k_i\}$.

2) Define the transition rate array $k = [k_1, \ldots, k_N]$. Then at a given simulation step, define elements of an accumulative reaction rate array as $\alpha_m = \sum_{i=1}^{m} k_i$.

3) Generate two random numbers $r_1$ and $r_2$ from a uniform distribution within [0,1]. The next time that an event will take place is given by $dt = \frac{1}{\alpha_N}\ln\left(\frac{1}{r_1}\right)$, so $t \to t + dt$, and the reaction channel taking place is the smallest integer $m$ satisfying $\alpha_m \geq r_2 \alpha_N$. If $s_m \neq 0$, update $s_m$ to 0. If $s_m = 0$, generate another random number $r_3$ from a uniform distribution within [0,1], update $s_m$ to -1 if $r_3 \leq (v'_{0\to -1}\bar{p}_2)/(v'_{0\to -1}\bar{p}_2 + v'_{0\to 1}\bar{p}_4)$, otherwise update $s_m$ to 1.

4) Repeat.

5) When it reaches the cell division time, for each nucleosome *i* generate a random number $r_4$ from a uniform distribution within [0,1]. If $r_4 \leq 0.5$ then $s_i = 0$, meaning that the histone is replaced by a nascent unmarked one; otherwise keep the original value of $s_i$, meaning the original histone is partitioned to this daughter cell being monitored. Here for simplicity we assume that the cell cycle time is fixed, which can be easily modified if variation of cell cycle time needs to be considered.

One can translate the above pseudo-code into any programming language such as Matlab, Python and C.

**5. Determine model parameters**

To perform the above numerical simulations, we need to determine the model parameters. A generally adopted strategy is to first determine or estimate the model parameters from experimental measurements.

Some parameter can be determined easily. If one assumes that some insulating elements constrain the histone modification patterns (43), one can estimate the number *N* from the DNA length within the constraints. For a gene length ~10k bp including the promoter regions, the nucleosome length *N* = 40. Without insulating elements, the model studies of Hodges and Crabtree show that an inherently bound histone pattern domain can be formed when the mark addition and removal enzymes have comparable catalytic activities (11). In that case the length of the domain is determined by the relative ratio between the addition and removal enzyme activities.

Below we discuss how to determine other parameters.

Nonspecific background free energy of binding of enzymes: Several experimental techniques, such as fluorescence recovery after photobleaching (FRAP) and fluorescence correlation spectroscopy (FCS), can provide quantitative information about protein-chromatin binding (25). In the literature what is usually reported is the fraction of enzymes bound to the histones. Below we discuss how to roughly estimate the free energy of binding from the data. Since these measurements are genome wide, therefore they reflect nonspecific protein-chromatin bindings instead of specific bindings facilitated by DNA-sequence specific elements.

Experimental data reveals that nonspecific protein-chromatin bindings are weak. Therefore we assume that the probability of having two neighboring nucleosomes occupied (from nonspecific background binding) at the same time is negligible. That is, for parameter estimation purpose we can neglect possible effects of the lateral interaction *J*, and treat each nucleosome as independent. Each histone can have two states: empty or occupied. Then respect to an arbitrary reference state with binding energy $\varepsilon_0$ and free enzyme concentration $c_0$, the binding energy with free enzyme concentration $c_{free}$ is $\varepsilon = \varepsilon_0 - \ln(c_{free}/c_0)$. From the Boltzmann distribution, the probability of observing a histone in the bound state is

$$p_H = \frac{\exp(-\varepsilon/k_B T)}{1+\exp(-\varepsilon/k_B T)}, \qquad \text{Eqn. 4}$$

then,

$$\varepsilon = -k_B T \ln \frac{p_H}{1-p_H}, \quad \varepsilon_0 = -k_B T \ln \frac{p_H c_0}{(1-p_H)c_{free}} \qquad \text{Eqn. 5}$$

From the cell volume and enzyme concentrations, we can estimate the total number of enzymes. Then from the measured fraction of bound enzymes, we obtain the total number of enzymes bound, noting that this number is also the total number of histones in the bound state. Next we can estimate the total number of nucleosomes from the genome size, assuming ~200 base pairs per nucleosome. The total number of (nucleosome) H3 proteins is twice the number of nucleosomes (since each nucleosome contains 2 copies of H3 proteins). From all these numbers we can estimate $p_H$.

**Insert Table 1**

Table 1 summarizes our estimations based on available experimental data, using 1 $\mu M$ as the reference free enzyme concentration $c_0$. Clearly our estimation is very rough. For example, we do not consider competition of binding from different types of enzymes. We also assume that every 200 base pairs form a nucleosome. This is clearly an overestimation of the total number of nucleosomes since there are nucleosome-free regions. Including these corrections reduces the number of free nucleosomes, and leads to a lower binding energy.

Notice that the estimated values of free energy of binding are positive. That is, nonspecific binding of enzymes on DNA is very weak at physiological histone and enzyme concentrations. Mechanistically this weak binding is reasonable. From the above table, the total number of nucleosomes is far more than that of the enzymes. That is, the number of substrates is much larger than the number of enzymes. Strong nonspecific binding would not allow a binding enzyme to move and interact with other nucleosomes, and seriously deplete the pool of free enzymes.

Free energy of binding of enzymes within the nucleation region: There is no quantitative information on the enzyme free energy of binding at specific genome region. The values are also affected by concentrations of the elements recruiting these enzymes. One piece of experimental information that can be used is the peaked distribution of the histone marks along the genome. The ratio between the peak value and that of the background value (for regions far away from the nucleation region) can be used to determine the specific binding affinities. That is, we require the ratio calculated from the model to match the experimental value (of Oct4 in the work of Zhang *et al.* (27)).

Enzyme lateral interactions: The values of $J_{\alpha\alpha}$ are chosen to reproduce the bell-like shaped histone methylation pattern centered around the nucleation region with a half-height width of about 10 nucleosomes, to represent the histone modification distribution pattern of Oct4 gene (11). In the work of Zhang *et al.* (27), for simplicity the same value of $J_{\alpha\alpha}$ is used for all enzymes. For interactions between different enzymes $J_{\alpha\beta}$ we simply assume that they may either be absent, or the enzyme interact unfavorably with several values examined to explore their effects on the epigenetic dynamics.

Enzyme rate constants: Without much direct experimental information, for simplicity we use the same rate constants for all four enzymes, and choose the value that reproduces the experimental observation that it takes about 5 cell cycles to switch Oct4 (11).

Histone exchange: The reported value of the histone exchange rate varies over a broad range and show cell-type dependence. In reality one may also expect dependence of histone exchange rate on the covalent marks. Active transcriptions can lead to higher histone exchange rate (38, 44), and thus different histone exchange rates may exist for euchromatins and heterochromatins. For simplicity though, Angel *et al.* uses a single value estimated from measurements on *Drosophila* cells (10). Zhang *et al.* adopt this value as well, and examine how changing the value affect the model behavior (27).

**Insert Figure 2**

6. **Perform computational studies**

Figure 2A shows a typical simulated trajectory using parameters roughly representing the gene Oct4. Clearly the *s* state of each nucleosome changes randomly and frequently. However, the system can exist in one collective epigenetic state, dominated by either repressive or active marks, for many cell cycles before stochastically switch to another state. A zoom-in of the trajectory (shown in Figure 2B) shows that a transition usually starts at one place, often within the nucleation region, then propagates outwards. Statistically the system still spends most of the time around either the repressive or active mark dominated states. That is, if one plots the fraction of time the system have *n* nucleosomes bearing repressive marks out of the *N* nucleosomes, one obtains a histogram with a bimodal distribution. In other words, the system exists as a bistable system.

Experimental studies reveal two essential molecular properties: enzymes can recognize the nucleosome marks and have mark-dependent free energy of binding, and enzymes bound to neighboring nucleosomes can interact laterally. Mathematically we use a quantity $\Delta\epsilon$ to reflect the mark-dependent free energy of binding, assuming that the binding energies for the addition or removal enzymes to a nucleosome bearing the corresponding (or antagonizing) mark are $\Delta\epsilon$ lower (or higher) than those binding to an unmodified nucleosome. That is, $\Delta\epsilon$ is an energetic penalty for mismatched binding between an enzyme and a nucleosome. The parameter $J_{\alpha\alpha}$ specifies the strength of lateral interactions between two neighboring enzymes of the same type. Figure 2C shows the calculated bistable region in the $\Delta\epsilon$ - $J_{\alpha\alpha}$ plane. Clearly a broad range of combinations of $\Delta\epsilon$ and $J_{\alpha\alpha}$ lead to bimodal distributions. A finite value of $J_{\alpha\alpha}$, with a critical minimum value ~2 $k_BT$, is necessary for generating bimodal distributions of the fraction of histones with repressive marks. Below this value of $J_{\alpha\alpha}$, increasing $\Delta\epsilon$ values does not lead to a bimodal distribution. The required value of $J_{\alpha\alpha}$ also increases sharply upon decreasing $\Delta\epsilon$. With $\Delta\epsilon \rightarrow 0$, the value of $J_{\alpha\alpha}$ needed for generating a bimodal distribution increases sharply. Within intermediate values, a decrease of $\Delta\epsilon$ can be compensated by an increase of $J_{\alpha\alpha}$. Therefore, these results demonstrate that $\Delta\epsilon$ and $J_{\alpha\alpha}$, representing the two observed molecular properties, are both sufficient and necessary to generate the epigenetic histone memory. This is an essential result and the working mechanism obtained from analyzing the CoPE model.

As mentioned above, a major and typical concern for modeling complex biological systems is that many parameters cannot be well determined experimentally. Therefore a key concept arising in quantitative biology studies is that if it holds for a broad range of model parameters, a mechanism is robust, and one has higher confidence that it reflects the true biology of the system; on the other hand, one should be skeptical and cautious on a mechanism that requires fine tuning model parameters. To show that the above-discussed physical mechanism is not a result of fine-tuning the model parameters, Zhang *et al.* performed simulations using 4096 sets of parameters in a 6-dimensional parameter space, with each dimension divided into 4 equally distributed grid points within a physically reasonable range. The 6 parameters are the free energy of binding and lateral interactions. They also used a *more* stringent criterion for the bistable region compared to what was used to generate Fig. 2D: clear separation between the epigenetic states with high and low average number of nucleosomes with repressive marks (>4.5), significant epigenetic memory with the average dwelling time on each epigenetic state > 2 cell cycles. It turns out that 1238 (30%) parameter sets satisfy the above requirement. Therefore, the mechanism is robust against parameter choices.

**Insert Figure 3**

**7. Identify insights from model studies and make testable predictions**

The above model simulations reveal a simple molecular mechanism for generating the epigenetic histone memory. Let's first consider an analogous situation. Suppose that there is a set of jigsaw puzzles (Figure 3A). A naughty kid randomly takes away pieces of the puzzle. You have two tasks:

1) Figure out what piece is missing. For more reliable reasoning the original pattern it is better to examine not only the slots of missing pieces, but also a larger region.

2) Put back a piece of puzzle the same as the missing one from a reservoir of spare puzzle pieces. The process should be faster than the process that the puzzle pieces are taken away. Otherwise quickly there would be accumulation of missing pieces, which make it more and more difficult for the reasoning in step 1.

Cells essentially have the same tasks, and the molecular properties of the involved molecular species ensure robust completeness of the tasks. Let's consider a collection of nucleosomes dominated by repressive marks (Figure 3B). After cell division, some of the nucleosomes are replaced by unmarked ones. The remaining nucleosomes with repressive marks preferentially recruit repressive mark enzymes relative to active mark enzymes---a "reading" process. Because of enzyme lateral interactions, these bound enzymes help the unmarked nucleosomes preferentially also recruit repressive mark enzymes, and add the repressive marks--- a "writing" process. Unlike genome inheritance, an epigenetic histone pattern, i.e., specific pattern of a given nucleosome, cannot be exactly inherited, but the overall pattern, repressive or active mark domination, can be rather faithfully maintained and inherited.

**Insert Table 2**

It may be easier to understand the above molecular mechanism using the two-nucleosome system. Suppose that originally both of the two nucleosomes bear repressive marks. After cell division, nucleosome 1 becomes unmarked. Table 2 gives the enzyme binding probabilities calculated from the Boltzmann distribution. The repressive-mark-bearing nucleosome 2 has higher probabilities of having the repressive mark addition or removal enzymes bound. Consequently, when it has these enzymes bound, nucleosome 1 also has higher probabilities of having the same enzyme bound. Overall nucleosome 1 has higher $\bar{p}_{\sigma_1=2}$ than $\bar{p}_{\sigma_1=4}$. That is, nucleosome 1 is more likely to add a repressive mark than an active mark to recover the original epigenetic pattern.

**Insert Figure 4**

An immediate conjecture from the above mechanism is that the system needs to reconstruct the epigenetic pattern faster than the perturbations coming from histone turnover, enzymatic reactions, and cell division. Indeed Fig. 4A shows that the model predicts sensitive dependence of the epigenetic state stability on the histone turnover rate *d*. Histone turnover is a major source of perturbations to the epigenetic pattern. A change of *d* value from 0.6 $h^{-1}$ to 1.2 $h^{-1}$ results in the average epigenetic state dwelling time changing from ~250 hours to 20 hours. Experimentally the value of *d* is difficult to measure accurately, and it varies over an order of magnitude (44-47). The value also depends on the cell types. Embryonic stem cells have a histone turnover rate higher than that of differentiated cells (47). It may be because that embryonic stem cells only exist transiently during the developmental process, and thus there is no selection pressure to maintain the epigenetic memory long. On the other hand, for cells like neurons, for which maintaining epigenetic information is crucial for their physiological functions, we predict that the value of *d* should be kept small. The model results in Fig. 4A also show that increasing the enzyme rate constant *v* can compensate an increased value of *d*. Increasing *v* allows faster recover of missed marks on nucleosome due to histone turnover. The 60-kDa HIV-Tat interactive protein (Tip60), a key member of the MYST family of histone acetyltransferases, can autoacetylate its lysine residue K327. Yang *et al.* report that a K327 deacetylated Tip60 only loses its catalytic activity by less one fold (48). However, Yuan *et al.* introduced this mutant into yeast and found this lack of autoacetylation is fatal for the survival of the organisms (49). Therefore, our model studies suggest that the enzyme activities (including concentrations) should be tightly regulated. Quantitative measurements are needed to test this prediction. If it is validated, then how are they robustly regulated?

Cell division is another major source of perturbations. Figure 4B shows that a mammalian cell is capable of quickly recovering (within a few hours) the original epigenetic pattern after losing about half of the histones due to cell division. A direct conjecture is that if one reduces cell cycle time so the cell has less time to recover from this perturbation, there is higher probability that the perturbation may accumulate over cell cycles and lead to faster switching of the epigenetic state. This conjecture is numerically proved by the results in Fig. 4C. This model result may help understand the experimental observation of Hanna *et al.* (50). These authors show that decreasing cell cycle time can accelerate the process reprogramming somatic cells to induced pluripotent stem cells. The result in Fig. 4C suggests that reduced cell cycle time may facilitate some genes to switch their epigenetic states and the cell could overcome the epigenetic barrier to achieve phenotypic transition.

## 8. Conclusion

Let's come back to the question we ask in the introduction. To understand how cells regulate and maintain phenotypes, a key step is to study how gene activities are regulated. Epigenetic histone modification is an essential part of the regulatory network. The mammalian cell reprogramming experiments reveal that epigenetic state switching is a rate-limiting step during the process (51-53). Recent advances in techniques such as CRISPR opens the possibility of easily editing the epigenome of a cell to artificially turn on or off selected genes. Therefore understanding the molecular mechanism(s) of epigenetic regulation is of both theoretical and practical importance.

In this chapter we use the CoPE model of epigenetic memory from Zhang *et al.* as an example to illustrate how one constructs and analyzes a mathematical model. We argue that even for a system with lots of unknowns, one can still perform certain level of mathematical modeling, and provide useful insights. One should be able to simplify and abstract the real system for modeling purpose, but in a well-controlled way so connections to the real physical quantities are transparent. Often there are a large number of model parameters that cannot be reliably constrained by available experimental data. One can still make qualitative and quantitative predictions through analyzing an ensemble of models with different parameter values. Last but not the least, modeling is not the end, but the starting point of another cycle of studies. While studying a complex biological system, modeling has its own strength and limitations, and an effort cohesively integrating modeling and experiments is always desirable.

The CoPE model should be viewed as an initial step to model the complex process of epigenetic regulation. In the above we mentioned a number of limitations of the model. In their review Rohlf *et al.* have a detailed discussion on the additional features future modeling efforts should take into account (18). For further development, more quantitative data and more molecular details would be needed. One may also adopt a multiscale modeling approach: using atomistic and coarse-grained modeling to explicitly include chromosome structure and provide inputs for the more coarse-grained modeling approach as described in this chapter.

Molecular dynamics (MD) simulation studies can provide down to atomistic level insight on some critical questions. For example, the histone turnover rate is an important parameter affecting the epigenetic dynamics. How is the rate affected by the histone covalent marks or DNA methylation? Does it show any sequence dependence? A single histone has many modifiable residues. How could different covalent marks crosstalk to each other through affecting binding affinities of various enzymes? In recent years, large-scale simulations based on structural details of biomolecules have advanced tremendously in terms of both simulation time scale (54) and simulation system size (55, 56). Since high-resolution crystal structures of nucleosome had been made available (57), systematic computational studies on the nucleosome histone modifications, starting from the atomistic level, become one of the important developments in the field of epigenetic research (58,59, 60). To support the atomistic scale simulation of the modified histone tails, commensurate efforts have been devoted to the force field developments to allow highly specific structural and energetic determination. For that purpose, *ab initio* quantum mechanics (QM) calculation, molecular mechanics (MM) or MD refinements, and experimental validation are all integrated (61). For example, a user-friendly and freely available platform for automated introduction of post-translocation modifications of choices to a protein 3D structure is presented by Vienna-PTM web server (http://vienna-ptm.univie.ac.at) (62). Furthermore, the *ab initio* QM/MM techniques have also been implemented to study histone modifying enzymes on their reaction mechanisms (63).

Besides the atomistic level approach, coarse-grained modeling from nucleosome toward chromatin level (64), with more or less structural basis and empirical interaction potentials, has also been developed accordingly. The type of models is quite adaptable to solve practical issues, without being restricted by time and spatial scales. For example, in a previous Monte-Carlo simulation of a 'mesoscale' chromatin model, histone tail flexibility, linker-histone electrostatic and orientation, magnesium ion induced electrostatic screening, and linker-DNA bending at physiological conditions, as well as thermal fluctuations and entropy effects are all considered (65). In a recent Brownian dynamics simulation study of DNA unrolling from the nucleosome, the mechanical forces from the histone core and effective electrostatic and site-specific binding of the DNA to the histone are considered, giving an estimation of the DNA-histone attraction at ~ 2.7 $k_B T$ per base pair (66). In another bead-spring model of chromatin, the flexible histone tails are made available for temporary electrostatic interaction with nucleosomes; the inter-nucleosomal interactions are thus mediated by the histone tails to allow distant communication in chromatin (67). A DNA lattice model in the framework of Ising-Markov approaches was developed as well, to describe transcription factor access to nucleosome DNA, taking into account intermediate protein binding state in which DNA is partially unwrapped from the histone octamer (68). Although these models cannot deal with the chemical nature of histone modification, they can be combined with atomistic or *ab initio* type of simulation studies to reveal how local histone modifications impact on global properties of nucleosome-nucleosome interactions and chromatic structures.

In summary, structure-based modeling efforts, both at atomistic and coarse-grained levels, will continue to help on analyzing existing experimental results, and guiding new experimental studies towards elucidating the molecular mechanism of epigenetic regulation and how it is coupled to other regulatory schemes such as transcription and translation.

References (<100)

1. Takahashi K, Yamanaka S. Induction of pluripotent stem cells from mouse embryonic and adult fibroblast cultures by defined factors. Cell. 2006;126(4):663-76.

2. Bannister AJ, Kouzarides T. Regulation of chromatin by histone modifications. Cell research. 2011;21(3):381-95.

3. Greer EL, Shi Y. Histone methylation: a dynamic mark in health, disease and inheritance. Nat Rev Genet. 2012;13(5):343-57.

4. Beisel C, Paro R. Silencing chromatin: comparing modes and mechanisms. Nat Rev Genet. 2011;12(2):123-35.

5. Black JC, Van Rechem C, Whetstine JR. Histone lysine methylation dynamics: establishment, regulation, and biological impact. Mol Cell. 2012;48(4):491-507.

6. Henikoff S, Shilatifard A. Histone modification: cause or cog? Trends in genetics : TIG. 2011;27(10):389-96.

7. Dodd IB, Micheelsen MA, Sneppen K, Thon G. Theoretical analysis of epigenetic cell memory by nucleosome modification. Cell. 2007;129(4):813-22.

8. Grewal SI, Klar AJ. Chromosomal inheritance of epigenetic states in fission yeast during mitosis and meiosis. Cell. 1996;86(1):95-101.

9. Thon G, Friis T. Epigenetic inheritance of transcriptional silencing and switching competence in fission yeast. Genetics. 1997;145(3):685-96.

10. Angel A, Song J, Dean C, Howard M. A Polycomb-based switch underlying quantitative epigenetic memory. Nature. 2011;476(7358):105-8.

11. Hodges C, Crabtree GR. Dynamics of inherently bounded histone modification domains. Proceedings of the National Academy of Sciences. 2012.

12. Hathaway Nathaniel A, Bell O, Hodges C, Miller Erik L, Neel Dana S, Crabtree Gerald R. Dynamics and Memory of Heterochromatin in Living Cells. Cell. 2012;149(7):1447-60.

13. Alsing AK, Sneppen K. Differentiation of developing olfactory neurons analysed in terms of coupled epigenetic landscapes. Nucleic acids research. 2013;41(9):4755-64.

14. Schwammle V, Jensen ON. A computational model for histone mark propagation reproduces the distribution of heterochromatin in different human cell types. PloS one. 2013;8(9):e73818.

15. Sedighi M, Sengupta AM. Epigenetic chromatin silencing: bistability and front propagation. Physical biology. 2007;4(4):246-55.

16. Dayarian A, Sengupta AM. Titration and hysteresis in epigenetic chromatin silencing. Physical biology. 2013;10(3).

17. Arnold C, Stadler PF, Prohaska SJ. Chromatin computation: Epigenetic inheritance as a pattern reconstruction problem. J Theor Biol. 2013;336:61-74.

18. Binder H, Steiner L, Przybilla J, Rohlf T, Prohaska S, Galle J. Transcriptional regulation by histone modifications: towards a theory of chromatin re-organization during stem cell differentiation. Physical biology. 2013;10(2):026006.

19. Schwab DJ, Bruinsma RF, Rudnick J, Widom J. Nucleosome switches. Physical review letters. 2008;100(22):228105.

20. Benecke A. Chromatin code, local non-equilibrium dynamics, and the emergence of transcription regulatory programs. Eur Phys J E. 2006;19(3):353-66.

21. Prohaska SJ, Stadler PF, Krakauer DC. Innovation in gene regulation: the case of chromatin computation. J Theor Biol. 2010;265(1):27-44.

22. David-Rus D, Mukhopadhyay S, Lebowitz JL, Sengupta AM. Inheritance of epigenetic chromatin silencing. J Theor Biol. 2009;258(1):112-20.

23. Raghavan K, Ruskin HJ, Perrin D, Goasmat F, Burns J. Computational micromodel for epigenetic mechanisms. PloS one. 2010;5(11):e14031.

24. Sontag LB, Lorincz MC, Georg Luebeck E. Dynamics, stability and inheritance of somatic DNA methylation imprints. J Theor Biol. 2006;242(4):890-9.

25. Steffen PA, Fonseca JP, Ringrose L. Epigenetics meets mathematics: towards a quantitative understanding of chromatin biology. BioEssays : news and reviews in molecular, cellular and developmental biology. 2012;34(10):901-13.

26. Rohlf T, Steiner L, Przybilla J, Prohaska S, Binder H, Galle J. Modeling the dynamic epigenome: from histone modifications towards self-organizing chromatin. Epigenomics. 2012;4(2):205-19.

27. Zhang H, Tian X-J, Mukhopadhyay A, Kim KS, Xing J. Statistical Mechanics Model for the Dynamics of Collective Epigenetic Histone Modification. Phys Rev Lett. 2014;112(6):068101.

28. Jacobs SA, Khorasanizadeh S. Structure of HP1 chromodomain bound to a lysine 9-methylated histone H3 tail. Science. 2002;295(5562):2080-3.

29. Margueron R, Justin N, Ohno K, Sharpe ML, Son J, Drury WJ, 3rd, et al. Role of the polycomb protein EED in the propagation of repressive histone marks. Nature. 2009;461(7265):762-7.

30. Kouzarides T. Chromatin Modifications and Their Function. Cell. 2007;128(4):693-705.

31. Cowieson NP, Partridge JF, Allshire RC, McLaughlin PJ. Dimerisation of a chromo shadow domain and distinctions from the chromodomain as revealed by structural analysis. Current biology : CB. 2000;10(9):517-25.

32. Canzio D, Chang EY, Shankar S, Kuchenbecker KM, Simon MD, Madhani HD, et al. Chromodomain-mediated oligomerization of HP1 suggests a nucleosome-bridging mechanism for heterochromatin assembly. Mol Cell. 2011;41(1):67-81.

33. So CW, Lin M, Ayton PM, Chen EH, Cleary ML. Dimerization contributes to oncogenic activation of MLL chimeras in acute leukemias. Cancer cell. 2003;4(2):99-110.

34. Buscaino A, Lejeune E, Audergon P, Hamilton G, Pidoux A, Allshire RC. Distinct roles for Sir2 and RNAi in centromeric heterochromatin nucleation, spreading and maintenance. Embo J. 2013.

35. Lin T, Ponn A, Hu X, Law BK, Lu J. Requirement of the histone demethylase LSD1 in Snai1-mediated transcriptional repression during epithelial-mesenchymal transition. Oncogene. 2010;29(35):4896-904.

36. Peinado H, Ballestar E, Esteller M, Cano A. Snail Mediates E-Cadherin Repression by the Recruitment of the Sin3A/Histone Deacetylase 1 (HDAC1)/HDAC2 Complex. Molecular and Cellular Biology. 2004;24(1):306-19.

37. Herranz N, Pasini D, Diaz VM, Franci C, Gutierrez A, Dave N, et al. Polycomb complex 2 is required for E-cadherin repression by the snail1 transcription factor. Mol Cell Biol. 2008;28(15):4772-81.

38. Gaffney DJ, McVicker G, Pai AA, Fondufe-Mittendorf YN, Lewellen N, Michelini K, et al. Controls of Nucleosome Positioning in the Human Genome. PLoS genetics. 2012;8(11):e1003036.

39. Mendenhall EM, Koche RP, Truong T, Zhou VW, Issac B, Chi AS, et al. GC-rich sequence elements recruit PRC2 in mammalian ES cells. PLoS genetics. 2010;6(12):e1001244.

40. Ku M, Koche RP, Rheinbay E, Mendenhall EM, Endoh M, Mikkelsen TS, et al. Genomewide analysis of PRC1 and PRC2 occupancy identifies two classes of bivalent domains. PLoS genetics. 2008;4(10):e1000242.

41. Xu C, Bian C, Lam R, Dong A, Min J. The structural basis for selective binding of non-methylated CpG islands by the CFP1 CXXC domain. Nature communications. 2011;2:227.

42. Consortium EP, Dunham I, Kundaje A, Aldred SF, Collins PJ, Davis CA, et al. An integrated encyclopedia of DNA elements in the human genome. Nature. 2012;489(7414):57-74.

43. Bushey AM, Dorman ER, Corces VG. Chromatin Insulators: Regulatory Mechanisms and Epigenetic Inheritance. Mol Cell. 2008;32(1):1-9.

44. Deal RB, Henikoff JG, Henikoff S. Genome-wide kinetics of nucleosome turnover determined by metabolic labeling of histones. Science. 2010;328(5982):1161-4.

45. Bhattacharya D, Talwar S, Mazumder A, Shivashankar GV. Spatio-temporal plasticity in chromatin organization in mouse cell differentiation and during Drosophila embryogenesis. Biophysical journal. 2009;96(9):3832-9.

46. Zee BM, Levin RS, Dimaggio PA, Garcia BA. Global turnover of histone post-translational modifications and variants in human cells. Epigenetics Chromatin. 2010;3(1):22.

47. Barth TK, Imhof A. Fast signals and slow marks: the dynamics of histone modifications. Trends in biochemical sciences. 2010;35(11):618-26.

48. Yang C, Wu J, Zheng YG. Function of the Active Site Lysine Autoacetylation in Tip60 Catalysis. PLoS ONE. 2012;7(3):e32886.

49. Yuan H, Rossetto D, Mellert H, Dang W, Srinivasan M, Johnson J, et al. MYST protein acetyltransferase activity requires active site lysine autoacetylation. EMBO J. 2012;31(1):58-70.

50. Hanna J, Saha K, Pando B, van Zon J, Lengner CJ, Creyghton MP, et al. Direct cell reprogramming is a stochastic process amenable to acceleration. Nature. 2009;462(7273):595-U63.

51. Pasque V, Jullien J, Miyamoto K, Halley-Stott RP, Gurdon JB. Epigenetic factors influencing resistance to nuclear reprogramming. Trends in genetics : TIG. 2011;27(12):516-25.

52. Ang YS, Gaspar-Maia A, Lemischka IR, Bernstein E. Stem cells and reprogramming: breaking the epigenetic barrier? Trends in pharmacological sciences. 2011;32(7):394-401.

53. Papp B, Plath K. Epigenetics of reprogramming to induced pluripotency. Cell. 2013;152(6):1324-43.

54. Piana S, Klepeis JL, Shaw DE. Assessing the accuracy of physical models used in protein-folding simulations: quantitative evidence from long molecular dynamics simulations. Current Opinion in Structural Biology. 2014;24(0):98-105.

55. Freddolino PL, Arkhipov AS, Larson SB, McPherson A, Schulten K. Molecular Dynamics Simulations of the Complete Satellite Tobacco Mosaic Virus. Structure. 2006;14(3):437-49.

56. Sanbonmatsu K, Blanchard S, Whitford P. Molecular Dynamics Simulations of the Ribosome. In: Dinman JD, editor. Biophysical approaches to translational control of gene expression. Biophysics for the Life Sciences. 1: Springer New York; 2013. p. 51-68.

57. Andrews A, Luge K. Nucleosome Structure(s) and Stability: Variations on a Theme. Annual Review of Biophysics. 2011;40:99-117.

58. Potoyan DA, Papoian GA. Regulation of the H4 tail binding and folding landscapes via Lys-16 acetylation. Proceedings of the National Academy of Sciences. 2012;109(44):17857-62.

59. Sanli D, Keskin O, Gursoy A, Erman B. Structural cooperativity in histone H3 tail modifications. Protein Science. 2011;20(12):1982-90.

60. Korolev N, Yu H, Lyubartsev AP, Nordenskiold L. Molecular Dynamics Simulations Demonstrate the Regulation of DNA-DNA Attraction by H4 Histone Tail Acetylations and Mutations. Biopolymers. 2014.

61. Grauffel C, Stote RH, Dejaegere A. Force field parameters for the simulation of modified histone tails. Journal of Computational Chemistry. 2010;31(13):2434-51.

62. Margreitter C, Petrov D, Zagrovic B. Vienna-PTM web server: a toolkit for MD simulations of protein post-translational modifications. Nucleic Acids Research. 2013;41(W1):W422-W6.

63. Zhang Y. Ab Initio Quantum Mechanical/Molecular Mechanical Studies of Histone Modifying Enzymes. In: York D, Lee T-S, editors. Multi-scale Quantum Models for Biocatalysis. Challenges and Advances in Computational Chemistry and Physics. 7: Springer Netherlands; 2009. p. 341-50.

64. Korolev N, Fan Y, Lyubartsev AP, Nordenskiöld L. Modelling chromatin structure and dynamics: status and prospects. Current Opinion in Structural Biology. 2012;22(2):151-9.

65. Grigoryev SA, Arya G, Correll S, Woodcock CL, Schlick T. Evidence for heteromorphic chromatin fibers from analysis of nucleosome interactions. Proceedings of the National Academy of Sciences. 2009.

66. Wocjan T, Klenin K, Langowski J. Brownian dynamics simulation of DNA unrolling from the nucleosome. The journal of physical chemistry B. 2009;113(9):2639-46.

67. Kulaeva OI, Zheng G, Polikanov YS, Colasanti AV, Clauvelin N, Mukhopadhyay S, et al. Internucleosomal interactions mediated by histone tails allow distant communication in chromatin. Journal of Biological Chemistry. 2012.

68. Teif VB, Ettig R, Rippe K. A Lattice Model for Transcription Factor Access to Nucleosomal DNA. Biophysical Journal. 2010;99(8):2597-607.

69. Fonseca JP, Steffen PA, Muller S, Lu J, Sawicka A, Seiser C, et al. In vivo Polycomb kinetics and mitotic chromatin binding distinguish stem cells from differentiated cells. Genes & development. 2012;26(8):857-71.

70. Muller KP, Erdel F, Caudron-Herger M, Marth C, Fodor BD, Richter M, et al. Multiscale analysis of dynamics and interactions of heterochromatin protein 1 by fluorescence fluctuation microscopy. Biophysical journal. 2009;97(11):2876-85.

**Glossary**

**List of Acronyms and Abbreviations**

S. Pombe: Schizosaccharomyces pombe

H3K4: Lysine 4 on histone H3

H3k9: Lysine 9 on histone H3

H3k27: Lysine 27 on histone H3

LSD1: lysine-specific demethylase 1

HDAC1: histone deacetylase 1

MLL1: Mixed Lineage Leukemia 1

PRC2 Polycomb Repressive Complex 2

FRAP: fluorescence recovery after photobleaching

FCS: fluorescence correlation spectroscopy

Tip60: HIV-Tat interactive protein

MD: molecular dynamics

PTM: post-translational modification

**Figure Legend**

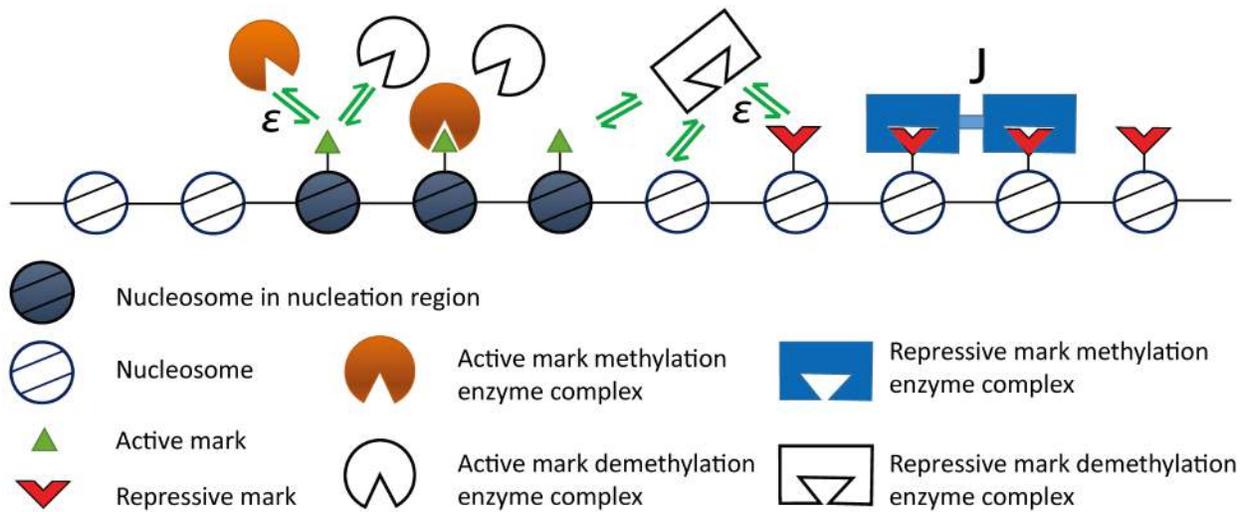

Figure 1. Schematic illustration of the CoPE model of Zhang *et al*. $\varepsilon$ denotes enzyme binding energy, $J$ denotes enzyme lateral interaction energy. Adapted from(27).

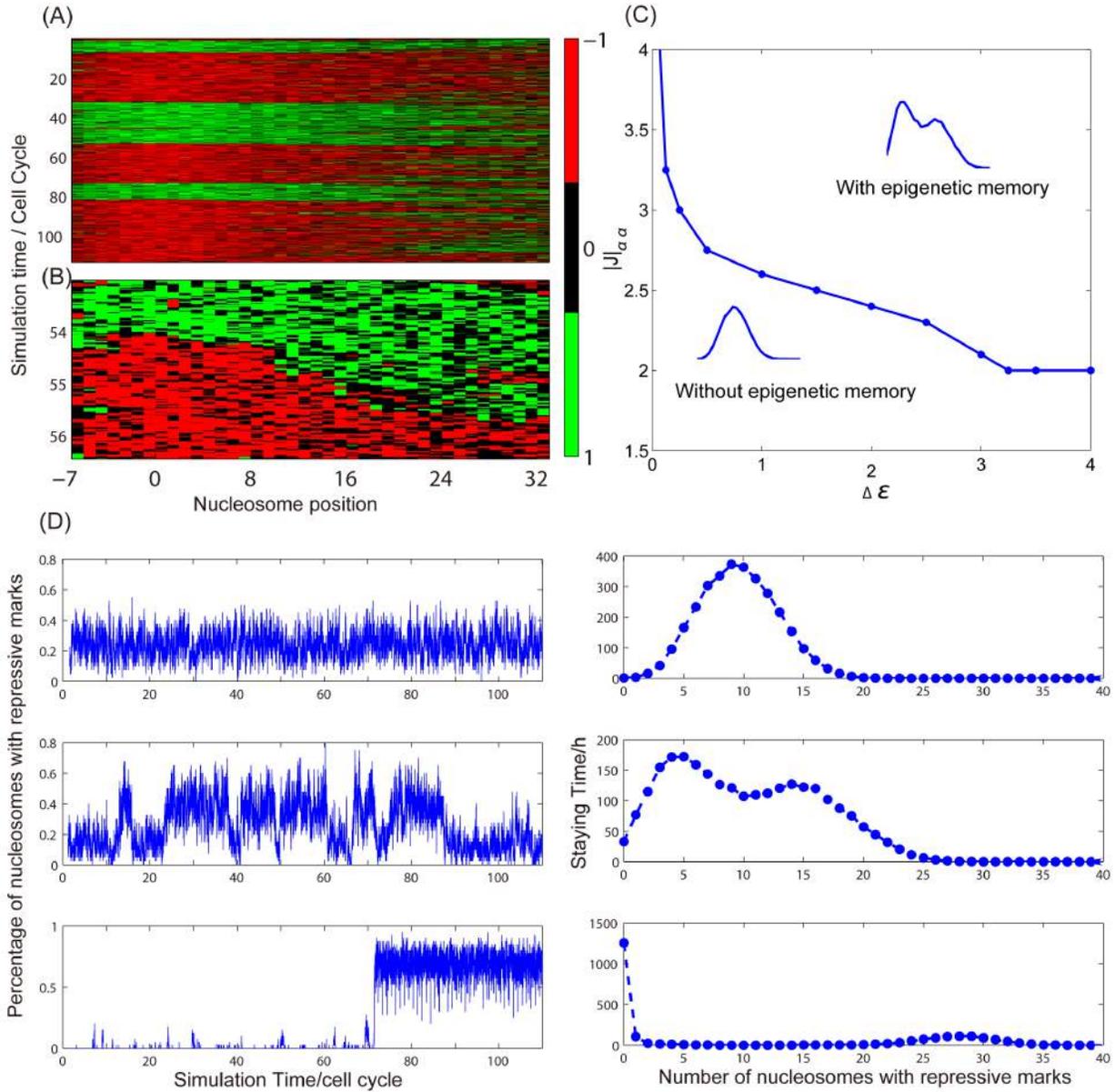

**Figure 2**. Simulation results using model parameters corresponding to Oct4. (A) Heat map representation of a typical simulation trajectory. (B) Zoom-in of the heat map in panel (A) showing epigenetic state transition. (C) Phase diagram on the $\Delta\varepsilon$ -$J$ plane to illustrate bistability mechanism. (D) Typical trajectories of the fraction of nucleosomes with repressive marks (left) and the corresponding probability distribution of observing given number of nucleosomes with repressive marks (right). All simulations are performed with $\Delta\varepsilon = 2$, but different $J_{\alpha\alpha}$ values, Upper panel: $J_{\alpha\alpha} = 0$, middle panel: $J_{\alpha\alpha} = 2:5$, lower panel: $J_{\alpha\alpha} = 3:5$. The dwelling time

distribution is obtained by averaging over 100 trajectories, each started with a randomly selected initial histone modification configuration, simulated for $10^3$ Gillespie steps, then followed by another $2 \times 10^3$ Gillespie steps for sampling. Adapted from (27).

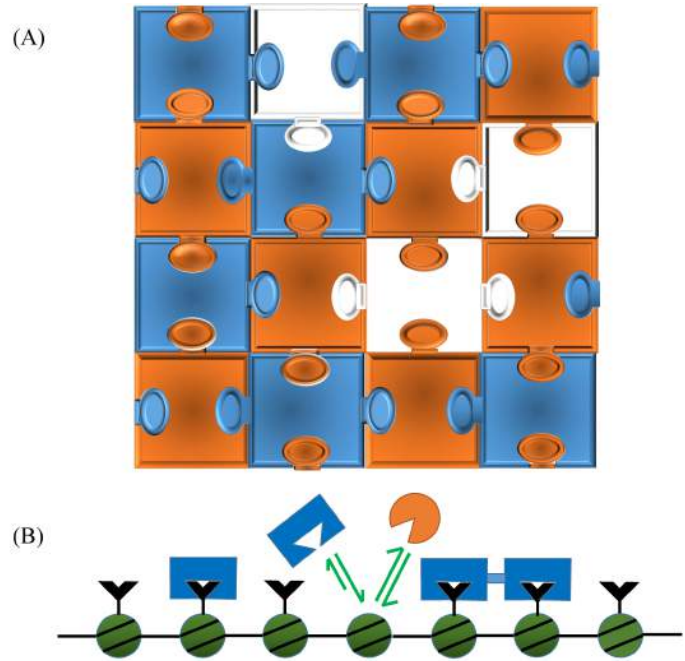

**Figure 3.** Schematic illustration of the reader/writer mechanism. (A) An analogous jigsaw puzzle reconstruction problem. (B) Reader-and-writer mechanism for epigenetic pattern reconstruction.

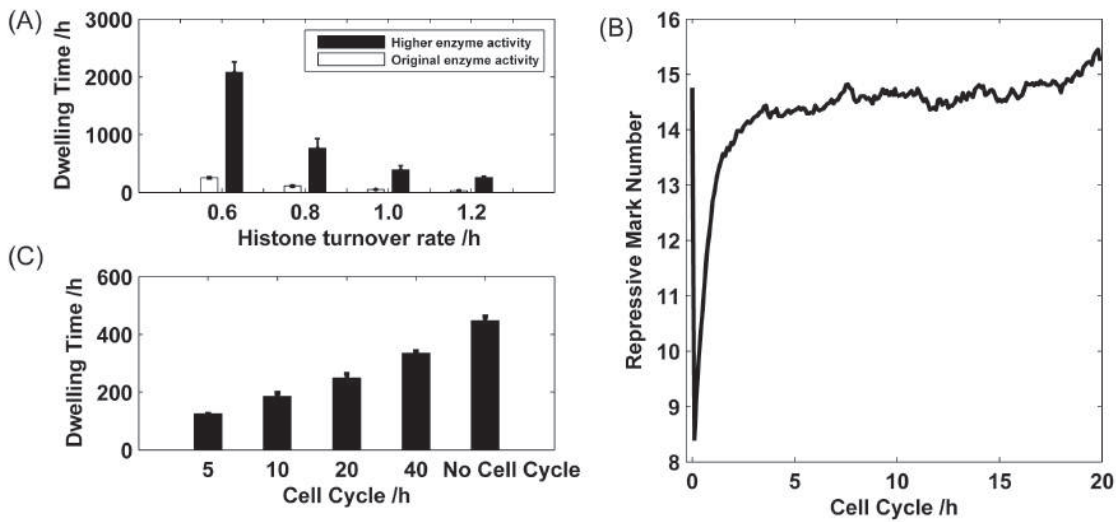

**Figure 4**. Different parameters affect histone epigenetic memory and dynamics. (A) Dependence of average dwelling time on histone exchange rate and different enzymatic activity. Higher enzymatic activity v=3.0, original enzymatic activity v=1.5. (B) After cell-cycle relaxation dynamics of the total number of nucleosomes with repressive marks. (C) Average state dwelling time as a function of the cell cycle. Adapted from (27).

Tables

|  | H3K9me3 | H3K27me3 | Refs |
|---|---|---|---|
| Enzyme | HP1α | Polycomb group (PcG) proteins | (25, 69, 70) |
| Cell source | Mouse L cells | Drosophila Neuroblasts / Embryo | (18,65,66) |
| Nuclear volume ($\mu M^3$) | 435 | 200 | (25) |
| Estimated nucleosome number | 21,120,000 (L cells) | 960,000 (Embryo (cycle 14)) | (25) |
| Nucleosome concentration | 80.6 mM (L cells) | 7.97 µM (Embryo (cycle 14)) | (25) |
| Measured enzyme bound fraction | 65% (Mouse NIH 3T3/iMEFs) | 18.93% (Drosophila Neuroblasts cells) | (69, 70) |
| Total enzyme concentration | 1µM | 380 nM (Drosophila Neuroblasts cells) | (69, 70) |
| Number of bound enzymes | 149477 | 10350 | Derived |
| $P_H$ | 0.004 | 0.0045 | Derived based on Eqn. (4) |
| $c_{free}$ | 0.35µM | 0.308µM | Derived |
| $\varepsilon$ | 4.5 $k_BT$ | 4.2 $k_BT$ | Derived based on Eqn. (5) |

**Table 1 Estimation of nonspecific binding energy from experimental data. Reproduced from (27).**

|  | σ1 = 1 | σ1 = 2 | σ1 = 3 | σ1 = 4 | σ1 = 5 |
|---|---|---|---|---|---|

| $\sigma_2 = 1$ | 0.285 | 0.0685 | 0.0685 | 0.010 | 0.010 |
| --- | --- | --- | --- | --- | --- |
| σ2 = 2 | 0.177 | 0.042 | 0.042 | 0.0062 | 0.0062 |
| σ2 = 3 | 0.177 | 0.042 | 0.042 | 0.0062 | 0.0062 |
| σ2 = 4 | 0.0032 | 0.0008 | 0.0008 | 0.0001 | 0.0001 |
| σ2 = 5 | 0.0032 | 0.0008 | 0.0008 | 0.0001 | 0.0001 |
| $\bar{p}_{\sigma_1}$ | 0.645 | 0.155 | 0.155 | 0.0227 | 0.0227 |

**Table 2 Calculated enzyme binding probabilities of a two-nucleosome system with $s_1 = 0$, and $s_2 = -1$. All model parameters are taken from Table 1 of Zhang *et al.* (27). Specifically, $J_{\alpha\alpha} = 3\ k_BT$, $\Delta\varepsilon = 2\ k_BT$.**